# Optimization and Performance of Bifacial Solar Modules: A Global Perspective

Xingshu Sun,[1] Mohammad Ryyan Khan,[1] Chris Deline,[2] and Muhammad Ashraful Alam[1,*]

[1] Network of Photovoltaic Technology, Purdue University, West Lafayette, IN, 47907, USA

[2] National Renewable Energy Laboratory, Golden, Colorado, 80401, USA

[*]Corresponding author: alam@purdue.edu

*Abstract* — With the rapidly growing interest in bifacial photovoltaics (PV), a worldwide map of their potential performance can help assess and accelerate the global deployment of this emerging technology. However, the existing literature only highlights optimized bifacial PV for a few geographic locations or develops worldwide performance maps for very specific configurations, such as the vertical installation. It is still difficult to translate these location- and configuration-specific conclusions to a general optimized performance of this technology. In this paper, we present a global study and optimization of bifacial solar modules using a rigorous and comprehensive modeling framework. Our results demonstrate that with a low albedo of 0.25, the bifacial gain of ground-mounted bifacial modules is less than 10% worldwide. However, increasing the albedo to 0.5 and elevating modules 1 m above the ground can boost the bifacial gain to 30%. Moreover, we derive a set of empirical design rules, which optimize bifacial solar modules across the world, and provide the groundwork for rapid assessment of the location-specific performance. We find that ground-mounted, vertical, east-west-facing bifacial modules will outperform their south-north-facing, optimally tilted counterparts by up to 15% below the latitude of 30º, for an albedo of 0.5. The relative energy output is reversed of this in latitudes above 30º. A detailed and systematic comparison with data from Asia, Africa, Europe, and North America validates the model presented in this paper. An online simulation tool (https://nanohub.org/tools/pub) based on the model developed in this paper is also available for a user to predict and optimize bifacial modules in any arbitrary location across the globe.

## I. Introduction

Solar photovoltaics (PV) has become one of the fastest growing renewable energy sources in the world as its cost has dropped dramatically in recent decades The present levelized cost of electricity (LCOE) of large-scale PV is already lower than that of fossil fuel in some cases [1]. New technological innovations will lower LCOE further. In this context, bifacial solar modules appear particularly compelling [2], [3]. In contrast to its monofacial counterpart, a bifacial solar module collects light from both the front and rear sides, allowing it to better use diffuse and albedo light, see Fig. 1(a). For example, Cuevas *et al.* [4] have demonstrated a bifacial gain up to 50% relative to identically oriented and tilted monofacial modules. Here, bifacial gain is defined as

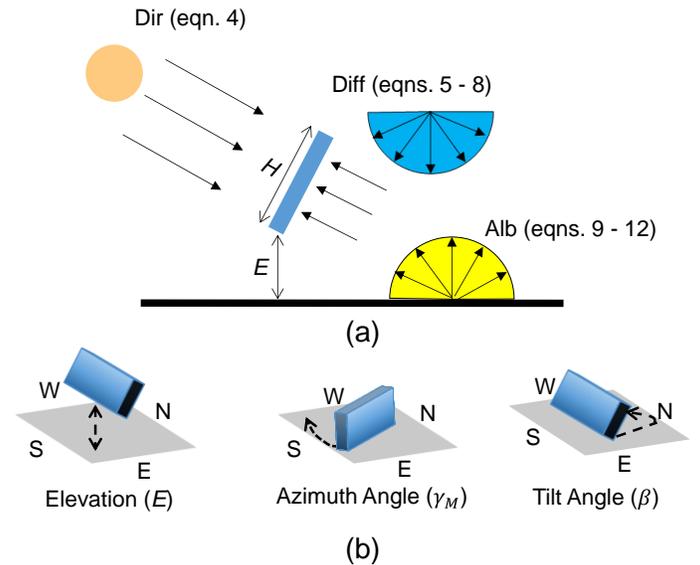

Fig. 1 (a) A schematic of a bifacial solar module with absorption of direct (Dir), diffuse (Diff), and ground-reflected albedo light (Alb). Equations used to calculate these irradiance components are labeled here. E and H denote the elevation and height (set to be 1 m in paper) of the solar module, respectively. (b) The three parameters discussed in this paper to optimize

$$\text{Bifacial Gain} = (Y_{Bi} - Y_{Mono})/Y_{Mono}, \quad (1)$$

where $Y_{Bi}$ and $Y_{Mono}$ are the electricity yields in kWh for bifacial and monofacial solar modules, respectively. Moreover, the glass-to-glass structure of bifacial modules improves the long-term durability compared to the traditional glass-to-backsheet monofacial modules. Also, many existing materialsthin-film PV technologies (e.g., dye-sensitized [5], CdTe[6], CIGS [7]) are readily convertible into bifacial solar modules. Due to the high efficiency and manufacturing compatibility into the bifacial configuration, silicon technologies, e.g., Si heterojunction cells, have received most attention [3]. This process compatibility, extra energy produced by the rear-side collection, reduced temperature coefficient, and longer module lifetime can potentially the installation cost as well as the LCOE significantly [8], [9]. This overall economic advantage persists despite that manufacturing bifacial solar modules can be more expensive than monofacial ones due to additional materials (e.g., dual glasses) and processes (e.g., screen-printing rear contacts). Based on these considerations, the International Technology Roadmap for Photovoltaic (ITRPV) anticipates the global market share of bifacial technology to expand from less than 5% in 2016 to 30% in 2027 [10].

The 50% bifacial gain for idealized standalone modules predicted by Cuevas *et al.* [4], however, is not always achievable in practice; thus, some of the highly optimistic projections regarding technology adoption may not be realistic. For example, intrinsic non-idealities, such as self-shading, can reduce the bifacial gain to less than 10% [11]. Therefore, one can accurately assess the performance potential and economic viability of bifacial modules only after accounting for the intrinsic non-ideal factors (e.g. self-shading) rigorously.

Toward this goal, several groups have reported on the performance of south-north-facing, optimally tilted, standalone bifacial solar modules—both numerically [11]–[14] and experimentally [15], [16]. These studies have shown that the deployment (e.g., elevation, orientation) and the environment conditions (e.g., irradiance intensity, ground albedo) dictate the energy output of bifacial solar modules, and the synergistic effects of these factors ought to be accounted for when evaluating the performance of bifacial technologies. Unfortunately, the analyses are confined to *only a few locations*, so these studies do not offer any guidance regarding the optimized configuration and the maximum energy output in a global context with very different irradiance and albedo.

Other groups have focused on worldwide analysis, but confined themselves to *specific configurations that are not necessarily optimal*. For example, Guo *et al.* [17] and *Ito et al.* [18] have presented worldwide analyses of east-west-facing, vertical bifacial solar modules. These vertical modules reduce soiling/snow losses [19], [20] and produce flatter energy output compared to their south-north-facing counterparts. Guo *et al.* concluded that for an arbitrary geographic location, an albedo threshold always exists above which *vertical* bifacial solar modules will outperform optimally tilted monofacial counterparts.

Apparently, location-specific, optimally tilted and oriented bifacial solar modules will produce even more energy than vertical modules. Indeed, the PV community will benefit greatly from a set of physics-based empirical equations that can calculate the optimum tilt and azimuth angles of bifacial solar modules given the geographic location, similar to those developed for monofacial ones [21]; however, such design guidelines are not currently available.

In this paper, we provide a global analysis and optimization of a variety of module configurations using our comprehensive opto-electro-thermal simulation framework. Our results reveal that the bifacial gain of ground-mounted bifacial modules is no more than ~*10%* across the globe for an albedo of 0.25, typical for groundcover of vegetation and soil. On the other hand, increasing albedo to 0.5 using artificial reflectors (e.g., white concrete) can double the bifacial gain to ~20%; further, elevating the module 1 m above the ground can improve the bifacial gain to ~30%. These results highlight the importance of highly reflective groundcover and module elevation for increasing/optimizing bifacial gain.

In addition, we will summarize our numerical results into a set of empirical equations that can analytically and optimally configure bifacial modules as a function of three design parameters—elevation (E), azimuth angle ($\gamma_M$), and tilt angle ($\beta$)—as schematically illustrated in Fig. 1(b). Our optimization results anticipate that for ground-mounted bifacial modules with an albedo of 0.5, east-west-facing verticallly installed bifacial modules ($Bi_{EW}$) will outperform south-north-facing optimally tilted ($Bi_{SN}$) ones by up to 15% for latitudes within 30º, and vice versa for latitudes above of 30º. In summary, our paper offers a global perspective on standalone bifacial solar modules to facilitate a more detailed LCOE calculation of this technology [22], [23].

The paper is organized as follows. Section II introduces the simulation framework. Section III presents the global performance of bifacial solar modules for various deployment scenarios. Section IV shows the derivation of a set of empirical equations that can analytically optimize bifacial solar modules for any arbitrary geographic location. Finally, Section V provides some concluding thoughts.

## II. SIMULATION FRAMEWORK

The modeling framework can be divided into three parts: 1) the geographic and temporal irradiance model integrated with the NASA meteorological database, 2) the geometric and analytical light-collection model, and 3) the physics-based electro-thermal-coupled model to calculate PV output power from solar insolations. Below, we discuss these three parts sequentially.

### A. Irradiance Model

**Solar Path.** First, we begin by calculating the position of the sun, i.e., the solar path, which is a prerequisite to obtaining the insolation and its collection by solar modules. In this paper, we use the NREL's solar position algorithm [24] implemented in the Sandia PV modeling library [25] to simulate the solar path—specifically, the solar zenith ($\theta_Z$) and azimuth ($\gamma_S$) angles at any arbitrary time and geographic location.

**Simulate GHI.** Next, we estimate the intensity of solar irradiance as follows. First, we calculate the irradiance intensity of global horizontal irradiance (GHI or $I_{GHI}$) on a minute-by-minute basis by inputting the solar path into the Haurwitz clear-sky model [26]–[28] implemented in PVLIB [25]. The clear-sky model assumes an idealized atmospheric condition (i.e., high irradiance transmission), which exists only for certain locations and weather conditions. Therefore, directly applying the clear-sky model often results in an overestimation of solar insolation. Fortunately, the NASA Surface Meteorology and Solar Energy database [29] provides open access to the satellite-derived 22-year *monthly* average insolation on a horizontal surface (kWh/(m$^2$day)), $In_{Hor}$, with a *spatial resolution* of $1 \times 1$ degree (latitude and longitude). The challenge here is that the database only provides monthly average irradiance, while accurate simulation of PV output necessitates a higher temporal resolution. Therefore, it is imperative to convert this monthly average into a minute-by-minute basis (given by the clear-sky model). To do so, we first assume constant daily horizontal insolation within a given month, and for each day thereof, we scale the minute-by-minute simulated $I_{GHI,clear-sky}$ to the average insolation $In_{Hor}$ to obtain the final $I_{GHI}$ following $I_{GHI} = I_{GHI,clear-sky} \times (In_{Hor}/\int I_{GHI,clear-sky}\,dt)$. Consequently, our approach allows us to simulate $I_{GHI}$ while fully accounting for the geographic and climatic factors.

**Irradiance Decomposition.** The calculated $I_{GHI}$ must be further decomposed into two segments: a) direct normal irradiance (DNI or $I_{DNI}$) and b) diffuse horizontal irradiance (DHI or $I_{DHI}$). The relationshp between these irradiance components can be expressed as follows:

$$I_{GHI} = I_{DNI} \times \cos(\theta_Z) + I_{DHI}. \quad (2)$$

Next, given the minute-by-minute sky clearness index $k_{T(M)}$, we use the Orgill and Hollands model [30] to empirically estimate $I_{DNI}$ and $I_{DHI}$ from $I_{GHI}$. Specifically, the clearness index is defined as the ratio between $I_{GHI}$ and extraterrestrial irrdiance ($I_0$) on a horiztonal surface, i.e., $k_{T(M)} = I_{GHI}/(I_0 \times \cos(\theta_Z))$, where $I_{GHI}$ is already known and $I_0$ can be analytically computed based on the day of year (DOY) [31], [32]. Inputting $k_{T(M)}$ into the Orgill and Hollands model, we can decompose $I_{GHI}$ into $I_{DNI}$ and $I_{DHI}$. An example of the simulated irradiances at Washington DC (38.9° N and 77.03° W) on June 10$^{th}$ is shown in Fig. 2. Other empirical models have also been proposed for GHI decomposition [33]–[35], but they produce comparable results [36]. The conclusions therefore are not affected by the model selection herein.

**Perez Model.** Next, we model the angular contributions of $I_{DHI}$ obtained earlier. Note that the angular distribution of $I_{DHI}$ is strongly correlated to the clearness index [21]. The diffuse irradiance that subtends the angular region adjacent to $I_{DNI}$ is referred as the circumsolar irradiance $I_{Diff(C)}$. $I_{Diff(C)}$ results from light scattering by aerosols particularly prevalent under clear sky. The diffuse irradiance that emerges from the Earth horizon at $\theta_Z = 90^o$ is called horizon brightening $I_{Diff(H)}$ and is caused by the Earth albedo irradiance. Both $I_{Diff(C)}$ and $I_{Diff(H)}$ are then superimposed on an isotropic diffuse irradiance background $I_{Diff(Iso)}$ to form an overall anisotropic diffuse irradiance spectrum [37]. The anisotropicity of the diffuse irradiance has a vital impact on the performance of solar modules due to the angularly dependent self-shading and light collection. Hence, we need to adopt the angle-dependent Perez model [37]–[38] obtained from [25] to decompose $I_{GHI}$ to correct for the overoptimistic estimation of PV energy production associated with a simpler isotropic model [21].

### B. Light-Collection Model

After calculating the irradiance, the second step involves calculating the integrated light collection by a solar module arising from each irradiance component, i.e., direct, diffuse, and albedo light, as depicted in Fig. 1(a). *In our view-factor based approach, we consider a single standalone module in two dimensions, which is equivalent to an infinitely long row of modules in three dimensions.*

**Direct Irradiance.** To evaluate the contribution of the direct irradiance $I_{DNI}$, we first need to know the angle of incidence (*AOI*) between $I_{DNI}$ and the front/rear surface of a solar module. Fortunately, *AOI* can be analytically calculated based on the solar $\theta_Z$ and $\gamma_S$ angles as well as the tilt ($\beta$) and azimuth ($\gamma_M$) angles of the solar module, expressed as

$$AOI = \cos^{-1}\{\cos(\theta_Z) \times \cos(\beta) + \sin(\theta_Z) \times \sin(\beta) \times \cos(\gamma_S - \gamma_M)\}. \quad (3)$$

For a bifacial solar module, the tilt $\beta_{Rear}$ and azimuth $\gamma_{M(Rear)}$ angles of the rear side are $(180^o - \beta_{Front})$ and $(\gamma_{M(Front)} + 180^o)$, respectively. Finally, the illumination by $I_{DNI}$ on both the front and rear sides of solar modules can be estimated as follows:

$$I_{PV:Dir(Front/Rear)} = (1 - R_{Loss}) \times \cos(AOI_{(Front/Rear)}) \times I_{DNI}, \quad (4)$$

where $R_{Loss}$ is the angle-dependent reflection loss from the module surface. Here, we use a widely applied empricial equation from [17], [38], [39], [40] that has demonstrated great accuracy in analytically approximating the angular reflectivity.

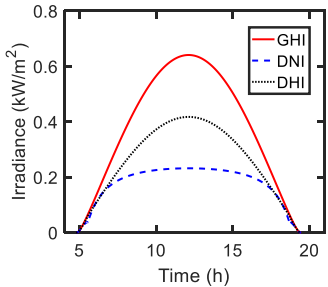

Fig. 2  Global horizontal irradiance (GHI), direct normal irradiance (DNI), diffuse horizontal irradiance (DHI) at Washington, DC (38.9° N and 77.03° W) on June 10$^{th}$.

**Diffuse Irradiance.** The calculation of diffuse light is more involved than that of direct light due to the anisotropic angular spectrum consisting of circumsolar, horizon brightening, and isotropic diffuse light. Each of the diffuse components requires a distinct approach to estimate its light collection by solar modules. A complete list of equations to calculate the contribution from diffuse light is given below,

$$I_{PV:Diff(Iso)} = (1 - R_{Loss}^{Int}) \times I_{Diff(Iso)} \times VF_{M \to Sky}, \quad (5)$$

$$I_{PV:Diff(C)} = (1 - R_{Loss}) \times I_{Diff(C)} \times \cos(AOI_{Cir}), \quad (6)$$

$$I_{PV:Diff(H)} = (1 - R_{Loss}) \times I_{Diff(H)} \times \sin(\theta_T), \quad (7)$$

$$I_{PV:Diff} = I_{PV:Diff(Iso)} + I_{PV:Diff(C)} + I_{PV:Diff(H)}, \quad (8)$$

where $VF_{M \to Sky} = (1 + \cos(\theta_T))/2$ is the module-to-sky view factor and $AOI_{Cir}$ is the angle of incidence for circumsolar diffuse light (equal to that of direct light until $\theta_Z > 85^0$). Note that because $I_{Diff(Iso)}$ is isotropic, rather than for one fixed angle, $R_{Loss}^{Int}$ in Eqn. (5) is the integral of reflection losses over the solid-angle window of the isotropic diffuse irradiance incident on the surface (see Eqns. (6a–6c) in [40]). Equations (5–8) enable us to analytically calculate the diffuse illumination on both the front and rear surfaces of solar modules.

**Albedo Irradiance.** Light-collection calculation of ground-reflected albedo light requires careful examination of self-shading, i.e., the ground shadow cast by solar modules, which substantially reduces illumination onto the ground, and consequently, the ground-reflected albedo irradiance both on the front and rear sides of a solar module [11], [41]. There are two categories of self-shading effects: 1) self-shaded direct and circumsolar diffuse irradiances, and 2) self-shaded isotropic diffuse irradiance, both of which are considered in our calculation as described below.

**Reflected Direct and Circumsolar Diffuse Irradiance.** As shown in Fig. 3(a), part of the ground does not receive $I_{Dir}$ and $I_{Diff(C)}$ due to self-shading by solar modules. Thus, only the unshaded portion of the ground can contribute to the reflected $I_{Dir}$ and $I_{Diff(C)}$ albedo light. It can be evaluated by

$$I_{PV(Front/Rear):Alb}^{DNI+Diff(C)} = (1 - R_{Loss}^{Int}) \times R_A \quad (9)$$
$$\times [I_{Dir} \times \cos(\theta_Z) + I_{Diff(C)} \times \cos(\theta_{Z(Cir)})]$$
$$\times \left[\frac{1 - \cos\left(\theta_{T\left(\frac{Front}{Rear}\right)}\right)}{2} - VF_{shaded \to Front/Rear} \times \frac{L_{Shadow}}{H}\right],$$

where $R_A$ is the ground albedo coefficient and the ground is assumed to be a Lambertian diffuse reflector, $\theta_{Z(Cir)}$ is the zenith angle of the circumsolar diffuse light (equals $\theta_Z$ until $\theta_Z > \theta_{Z(Max)} = 85^o$), $L_{Shadow}$ is the length of the shadow cast by the solar module, $H = 1$ m is the module height, and $VF_{shaded \to Frpmt/Rear}$ is the view factor from the shaded region of the ground to the front/rear side. We calculate $L_{Shadow}$ and $VF_{shaded \to Front/Rear}$ analytically following the methodologies in [42]–[44].

**Reflected Isotropic Diffuse Irradiance.** Blocked by solar modules, only a fraction of isotropic diffuse irradiance from the sky can reach to the ground and be reflected, see Fig. 3(b). Self-shading due to sky masking of $I_{Diff(Iso)}$ erodes the albedo collection of solar modules, because $I_{Diff(Iso)}$ depends strongly on the location of the ground ($x$) from which the view factor $VF_{x \to sky}(x)$ is calculated [38], i.e.,

$$VF_{x \to sky}(x) = 1 - (cos(\theta_1) + cos(\theta_2))/2. \quad (10)$$

The masking angles $\theta_1$ and $\theta_2$ at position $x$ are labeled in Fig. 3(b). Note that only a portion of the reflected $I_{Diff(Iso)}$ illuminates the rear side of a solar module, characterized by the view factor from position $x$ to the front/rear side, $VF_{x \to Front/Rear}(x) = 1 - VF_{x \to Sky}(x)$. Finally, one must integrate the albedo irradiance collected by the solar module over the ground to estimate the total illumination

$$I_{PV(Front/Rear):Alb}^{Diff(Iso)} = 1/H \times (1 - R_{Loss}^{Int}) \times R_A \times \quad (11)$$
$$I_{Diff(Iso)} \times \int_{-\infty}^{+\infty} VF_{x \to sky}(x) \times VF_{x \to Front/Rear}(x) \, dx.$$

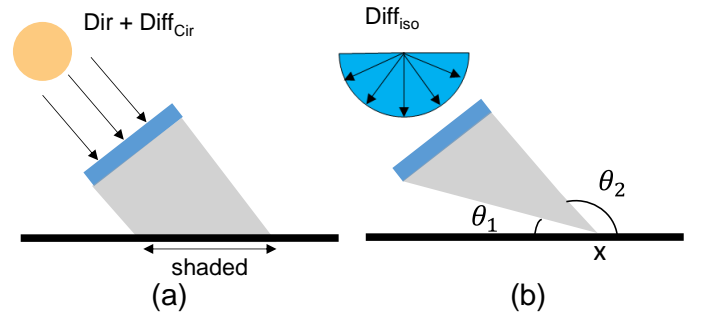

Fig. 3. Self-shading of albedo light from (a) direct (Dir) and circumsolar diffuse light (Diff_cir) and (b) isotropic diffuse light (Diff_iso).

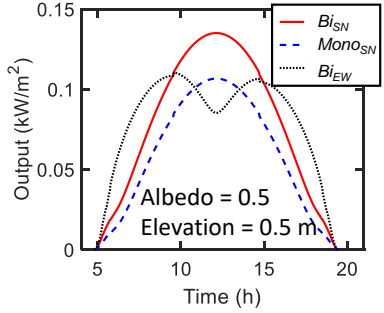

Fig. 4  Electricity output of a solar module in three configurations on a minute-to-minute basis at Washington, DC (38.9° N and 77.03° W) on June 10th.

Here, Eqn. (11) assumes an infinitely large ground reflector, which yields slightly higher albedo light compared to the finite ground reflector used in [11], [14]. Obviously, our framework is general and can account for finite ground correction, if needed.

Eventually, the total contribution of the albedo irradiance on the front/rear side is given by the sum of Eqns. (9–11):

$$I_{PV(Front/Rear):Alb} = I^{Diff(Iso)}_{PV(Front/Rear):Alb} + I^{DNI+Diff(C)}_{PV(Front/Rear):Alb}. \quad (12)$$

The light-collection model enables us to rigorously calculate the total illumination on both the front and rear sides of a bifacial solar module by accounting for self-shaded albedo light. Knowing the total amount of light incident on the module, we next couple this optical illumination to the electro-thermal model of the module to assess the total energy production by a bifacial solar module. This crucial aspect of the calculation has sometimes been omitted in various publications [44], [45].

### C.  Electro-Thermal Module Model

**Power Conversion Efficiency.** In the third and final step of the overall model, we must convert the incident light into electrical output. In this paper, the energy conversion from solar illumination into electricity is estimated as follows:

Table. 1 Modeling Framework Validation Against Literature

| Location (Type) | Elevation / Module Height (m) | Albedo / Bifaciality | Tilt Angle / Facing | Reported Bifacial Gain (%) | Calculated Bifacial Gain (%) | Difference (%) |
|---|---|---|---|---|---|---|
| Cairo (Sim.) [11] | 1 / 0.93 | 0.2 / 0.8 | 26° / South | 11.0 | 11.1 | -0.1 |
| Cairo (Sim.) [11] | 1 / 0.93 | 0.5 / 0.8 | 22° / South | 24.8 | 25 | -0.2 |
| Oslo (Sim.) [11] | 0.5 / 0.93 | 0.2 / 0.8 | 51° / South | 10.4 | 13.6 | -3.2 |
| Oslo (Sim.) [11] | 0.5 / 0.93 | 0.2 / 0.8 | 47° / South | 16.4 | 22.8 | -6.4 |
| Hokkaido* (Exp.) [46] | 0.5 / 1.66 | 0.2 / 0.95 | 35° / South | 23.3 | 25.7 | -2.4 |
| Hokkaido* (Exp.) [46] | 0.5 / 1.66 | 0.5 / 0.95 | 35° / South | 8.6 | 13 | -4.4 |
| Albuquerque (Exp.) [16] | 1.08 / 0.984 | 0.55 / 0.9 | 15° / South | 32.5** | 30.2 | 2.3 |
| Albuquerque (Exp.) [16] | 1.08 / 0.984 | 0.55 / 0.9 | 15° / West | 39** | 36.7 | 2.3 |
| Albuquerque (Exp.) [16] | 1.03 / 0.984 | 0.25 / 0.9 | 30° / South | 19** | 14.6 | 4.4 |
| Albuquerque*** (Exp.) [16] | 0.89 / 0.984 | 0.25 / 0.9 | 90° / South | 30.5** | 32.2 | -1.6 |
| Golden (Exp.) **** | 1.02 / 1.02 | 0.2 / 0.6 | 30° / South | 8.3 | 8.6 | -0.3 |
| * Only data from May to August were used to eliminate snowing effects. | | | | | | |
| ** Average bifacial gain of multiple test modules was used. | | | | | | |
| *** The east-west-facing vertical modules measurement in [16] shows great discrepancy between two modules; therefor, it is not included here. | | | | | | |
| **** Bifacial measurement (12/2016 to 08/2017) performed by the National Renewable Energy Laboratory. | | | | | | |

$$P_{PV} = I_{PV(Front)} \times \eta_{Front} + I_{PV(Rear)} \times \eta_{Rear}, \quad (13)$$

where $P_{PV}$ is total output power by bifacial solar modules, $\eta_{Front}$ and $\eta_{Rear}$ are the front- and rear-side efficiencies, respectively, and $I_{PV(Front)}$ and $I_{PV(Rear)}$ denote the front- and rear-side illumination of solar modules, respectively. Although the model is general and can be used for any technology, for illustration, we use the performance parameters obtained from commercially available bifacial solar module Bi60 manufactured by Prism Solar [47]. Specifically, the standard test condition (STC) efficiency of the front side for the simulated bifacial module $\eta_{Front(STC)} = 17.4\%$. The bifaciality of the module, which is defined as the ratio between the rear-side and front-side efficiencies, is $\eta_{Rear(STC)}/\eta_{Front(STC)} = 90\%$, corresponding to $\eta_{Rear(STC)} = 15.6\%$.

**Electro-Thermal Model.** The efficiency ($\eta(T_M)$) of bifacial solar modules in the field also depends on the real-time operating temperature described by

$$\eta(T_M) = \eta_{(STC)} \times \{1 + \beta \times (T_M - 298 \text{ K})\}. \quad (14)$$

Here, $\beta = -0.41\%/K$ is the temperature coefficienct retreived from [47] and $T_M$ is the module temperature. Under solar illumination, self-heating elevates $T_M$ above the ambient temperature $T_A$ [48]. Due to the additional rear-side absorption relative to monofacial, the bifacial module is expected to have greater energy input. However, bifacial modules are more transparent to sub-bandgap photons than monofacial modules, resulting in less self-heating [49]. Indeed one can solve the energy-balance equation self-consistently to obtain $T_M$, but this approach is only amenable to numerical methods and is not ideal for large-scale simulation. Hence, we use an analytical formula developed by Faiman [50] that empirically calculates $T_M$ based on the illumination and windspeed as follows:

$$T_M = T_A + \frac{I_{PV(Front)} + I_{PV(Rear)}}{U_0 + U_1 \times WS}, \quad (15)$$

where $WS$ denotes the wind speed that dictates convective cooling; and $U_0$ and $U_1$ are empirical fitting parameters contingent on module type and deployment (e.g., open rack and rooftop). Equation 15 calculates the module temperature based on both the front and rear solar absorption, thereby has explicitly considered temperature variation due to different ground albedo (e.g. vegetation vs. concrete). In this paper, we calibrate $U_0$ and $U_1$ to the nominal operating cell temperature ($NOCT = 47C^o$) of the Prism Solar Bi60 bifacial solar modules [47]. Global monthly average windspeed and ambient temperature data sets, also provided by the NASA meteorological database [29], are used in (15) to calculate $T_M$, and sequentially, the temperature-corrected efficiency of bifacial solar modules in this paper. Note that the windspeed and $T_A$ data in [29] involve monthly averages; in other words, our simulation assumes that the windspeed and $T_A$ are constant within a month. For locations with high intra-day temperature variation, the results may overestimate the energy yield since the highest diurnal temperature (when solar modules generate most power) can be higher than the average (a morning-to-noon temperature difference up to 45 ºC in desert environments [51]) and therefore significantly reduces the efficiency. Accounting for the hourly variation of $T_A$ and windspeed will improve the accuracy of the results, which will be an important aspect of future research on the topic.

**Power Loss due to Nonuniform Illumination.** As demonstrated by both simulation and experiments, self-shading can cause spatially nonuniform illumination on the rear surface of solar modules [11], [52], [53]. Equation (13) neglects this additional power loss from non-uniform illumination distribution. Note that elevating modules above the ground improves the illumination uniformity and reduces the loss associated with nonuniform illumination. Furthermore, the homogeneous front-side illumination can also offset the nonuniformity at the rear side and mitigate the corresponding loss. Nonetheless, if needed, the inclusion of such performance degradation can be easily achieved in our framework by using the analytical method described in [54].

*D. Simulation Demonstration*

To validate the aforementioned comprehensive simulation framework, we benchmark our results against the available data from the literature (including simulation [11] and experiments [16], [46]) as well as field data measured by the National Renewable Energy Laboratory. Among these data sets, Ref. [11] performed sophisticated ray-tracing simulation for optimizing annual production of bifacial modules in two different locations (i.e., Cairo and Oslo). Sugibuchi *et al*. [46] measured bifacial gain with two different albedo coefficients (grass versus shell grit), and here only data from May to August is used to eliminate snowing effects. The measured data from Ref.[16] was taken at Albuquerque, New Mexico, by the Sandia National Labororaties from 02/2016 to 02/2017, and it covers variously configured bifacial modules (e.g., 15º tilted east-west and 30º tilted south-north facing bifacial modules). Finally, the field data recorded by NREL were taken at Golden, Colorado, dating from 12/2016 to 08/2017. Note that the geographic locations of our benchmark results span across Asia, Africa, Europe, and North America.

Remarkably, our results match the bifacial gain reported in the literature within 6.4%. This excellent match was obtained even though our framework uses the NASA 22-year average meteorological database and assumes idealities such as infinite-size ground reflectors and obstruction-free shading. The benchmark results against field measurement are summarized in Table 1. The framework allows us to simulate and optimize the performance of standalone bifacial solar modules with different configurations (e.g., bifaciality, orientation, elevation, albedo) at any arbitrary time and geographic location. For example, Fig. 4 summarizes the simulated output power of three unique types of solar modules: 1) south-north-facing monofacial ($Mono_{SN}$), 2) south-north-facing bifacial ($Bi_{SN}$), and 3) east-west-facing bifacial ($Bi_{EW}$). These modules are all elevated 0.5 m above the ground with an albedo of 0.5 typical for white concrete. $Bi_{EW}$ is tilted 90°, i.e., vertical installation, whereas the tilt angles of $Mono_{SN}$ and $Bi_{SN}$ are optimized (for maximum production) at 37° and 48°, respectively.

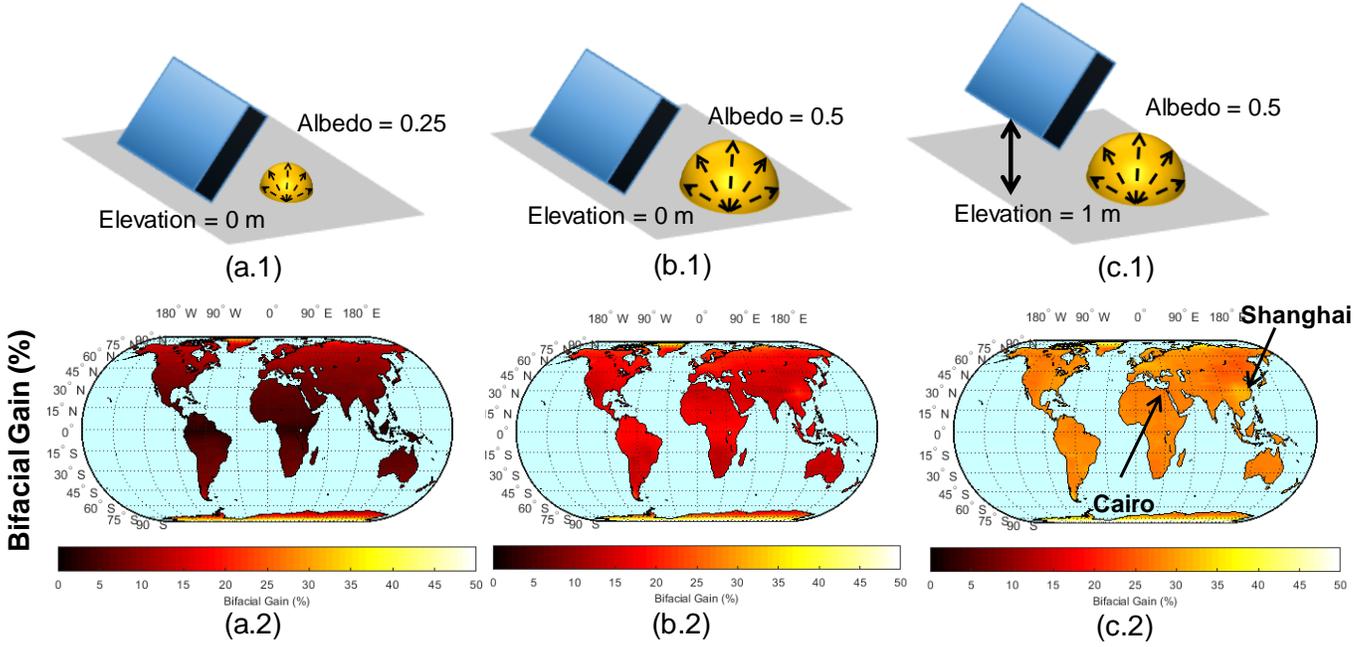

Fig. 5  Three different deployment scenarios of bifacial solar modules are simulated (depicted in the 1st row), i.e., (a) ground mounted with a ground albedo of 0.25, (b) ground mounted with a ground albedo of 0.5, and (c) 1m elevated with a ground albedo of 0.5. Global maps of these scenarios showing optimal bifacial gain (2nd row).

In the following section, we will extend our single-day analysis to the annual performance of differently configured solar modules in a global context, while fully exploring the effects of self-shading on the performance and optimization of bifacial solar modules. We note that our framework has been made available in an online simulation tool, i.e., Purdue University Bifacial Module Calculator [55]. The tool is capable of simulating and optimizing the performance of bifacial modules worldwide.

III. GLOBAL PERFORMANCE OF BIFACIAL SOLAR MODULES: A SUMMARY OF THE KEY RESULTS

Global maps of location- and configuration-specific performance of optimized bifacial solar modules have not been reported in the literature. Hence, we apply the rigorous framework presented in Sec. II to investigate bifacial gain of bifacial modules relative to their monofacial counterpart worldwide. For concreteness, we will focus on the worldwide results for three cases: (a) ground mounting with a ground albedo of 0.25 (natural ground reflector such as vegetation and soil), (b) ground mounting with a ground albedo of 0.5 (white concrete), and (c) 1 m elevation with a ground albedo of 0.5. *Here, we will illustrate that only limited bifacial gain is achievable across the entire world due to the low albedo of natural groundcover and self-shading of albedo light; however, one can substantially improve the bifacial gain by deploying highly reflective groundcovers and elevating the modules above the ground to reduce self-shading.* For a comprehensive comparison of bifacial performance, the supplementary material includes an extensive table of global maps of optimal deployment, bifacial gain, and annual electricity production for a broad range of elevation and ground albedo.

**Bifacial Gain.** Figure 5(a.2) shows that an albedo of 0.25 (typical for natural groundcover) results in a bifacial gain of less than 10% globally, even when the *ground-mounted* bifacial solar modules have been optimized for azimuth and tilt angles to maximize annual energy production. The limited bifacial gain herein is attributable to 1) the low ground albedo coefficient as well as 2) performance erosion due to self-shading. To further improve the bifacial gain, one must either increase the ground albedo coefficient, elevate modules above ground to reduce self-shading, or apply both simultaneously. Indeed, our results elucidate that increasing the ground albedo to 0.5 can boost the bifacial gain of ground-mounted modules to ~20% globally, as shown in Fig. 5(c.2). The substantial improvement of bifacial gain encourages the development of cost-effective artificial ground reflectors to supersede natural groundcovers. In addition, our simulation also predicts that elevating the module 1 m above the ground can further increase the bifacial gain to ~30% by recovering self-shading induced losses, see Fig. 5(b.2). However, elevating modules can result in additional installation cost; so, careful optimization of module elevation is required to maximize the bifacial gain while restraining installation cost. In the next section, we will derive a set of empirical rules to calculate the optimum elevation analytically.

**Clearness Index.** The performance of bifacial solar modules also depends on the local climatic condition, i.e., the annual sky clearness index $k_{T(A)}$, which indicates the amount of extraterrestrial irradiance transmitting through the atmosphere and reaching to the ground. Interestingly, bifacial gain decreases with clearness index, i.e., the absolute bifacial gain is ~5% higher in Shanghai than Cairo as shown in Fig. 5 (c.2). This increase in the bifacial gain is due to the higher concentration of diffuse light in the lower-transmitting atmosphere in Shanghai ($k_{T(A)} \approx 0.35$ in Shanghai compared to $k_{T(A)} \approx 0.7$ in Cairo). *Therefore, despite the lower total solar insolation, bifacial solar modules benefit more in Shanghai than Cairo due to the additional rear-side absorption of diffuse light.* This finding—i.e., bifacial modules are more advantageous in cloudier locations—has a profound yet practical implication on the adoption of bifacial modules globally. Note that the analytical equations developed to estimate bifacial gain in [56]–[58] do not always account for the clearness index, so the results may not be accurate. Hence, great caution should be taken when applying these equations to evaluate the location-specific performance of bifacial solar modules.

In this section, we have summarized our key results that for ground-mounted modules with an albedo of 0.25, the bifacial gain of fully optimized bifacial modules is less than 10% worldwide. Increasing the albedo to 0.5 and elevating modules 1 m above the ground, one can increase the bifacial gain up to ~30% globally. In the following section, we will explain how these optimizations were achieved and present a set of empirical guidelines for deploying bifacial modules.

## IV. WORLDWIDE OPTIMIZATION OF BIFACIAL SOLAR MODULES: PHYSICS AND METHODOLOGY

As already highlighted, there are three design parameters to optimize the electricity yield of bifacial modules—elevation ($E$), azimuth angle ($\gamma_M$), and tilt angle ($\beta$). These parameters are mutually dependent; specifically, optimal azimuth and tilt angles are a function of elevation. To isolate the mutual correlation among these parameters, we optimize the energy yield of bifacial modules by changing a single parameter, while keeping the other two parameters constrained. In this section, we specifically discuss the 1) minimum elevation $E_{95}$ to achieve 95% of maximum energy production; 2) optimum azimuth angle at fixed elevation, 3) finally, optimum tilt angle for given $E$ and $\gamma_M$. More importantly, for each parameter, we have derived a set of empirical equations that can analytically estimate the optimal value for an arbitrary location.

### A. Elevation

**Effect of Elevation.** An important factor affecting the performance of bifacial modules is their elevation above the ground. Highly elevated modules suffer considerably less from self-shading as shown in [11], [14], [41], which accords with our results in Sec. III. Therefore, elevation is a crucial design parameter to optimize the performance of bifacial solar modules. However, as the elevation continues to increase, the loss due to self-shading diminishes gradually until its effect is completely negligible. Hence, for infinitely large ground reflectors, the energy production of bifacial modules plateaus at high elevation above the ground [11], [41] and elevating the module further does not improve energy yield, see Fig. 6(a).

The elevation cutoff where production of bifacial solar modules starts to saturate is valuable to installers for minimizing the installation cost while preserving sufficient electricity yield. So, we estimate the average minimum elevation ($E_{95}$) to achieve 95% of the maximum energy production (i.e., self-shading free) as a function of latitude at a fixed ground albedo, see Fig. 6(b). It is noteworthy that $E_{95}$ decreases almost linearly with latitude, which is attributable to the suppressed self-shading by higher optimal tilt angle at higher latitude. In addition, $E_{95}$ rises with higher ground albedo up to almost 3 m near the Equator. Higher ground albedo increases the contribution of albedo light, making bifacial modules more susceptible to self-shading. Thus, $E_{95}$ has to increase to compensate the added self-shading loss.

**Empirical Equations.** By applying linear regression to the results in Fig. 6, we derive a set of empirical equations to estimate $E_{95}$ as a function of module height, latitude, and ground albedo, see Eqns. (A1–A2) in Table A1 of the appendix. The relative error of the empirical equations

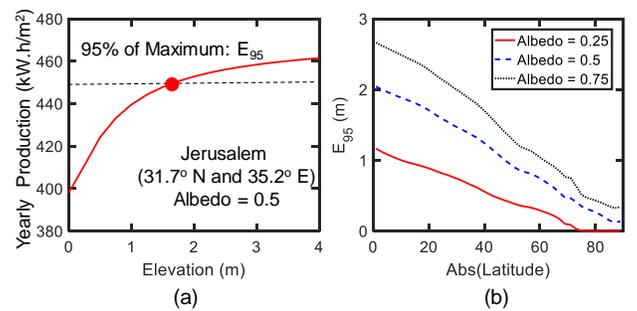

Fig. 6 (a) Yearly electricity production of optimally oriented and tilted bifacial solar modules with a height of 1 m as a function of elevation at Jerusalem (31.7° N and 35.2° E). The ground albedo is 0.5. The dashed line is the cutoff for 95% of the self-shading-absent maximum energy yield and red circle is the minimum elevation $E_{95}$ to achieve this threshold. (b) $E_{95}$ of bifacial solar as a function of absolute latitude for ground albedos of 0.25, 0.5, and 0.75. Note that the minimum elevation for each latitude in this plot is the average over longitudes with different clearness indexes.

compared to our numerical results is less than 1% for realistic albedo coefficients (from 0.25 to 0.75). Hence, these equations can assist installers to minimize the installation cost of elevating modules without sacrificing energy production. Note that Eqns. (A1–A2) assume a large ground reflector area (> 100 times the module area [11]); otherwise, $E_{95}$ is expected to drop because of the reduced view factor between the small ground area and the bifacial modules at high elevation.

Note that elevating solar modules can also enhance convective cooling power (wind speed increases with elevation [59]), thereby reduce the operating temperature. This cooling gain can boost the efficiency as well as improve the long-term durability of solar modules [48]. On the hand other, it must be pointed out again that elevating modules above the ground can impose additional installation expenditure (contingent on labor and material cost), but the empirical rules derived here does not account for these additional costs. Thus, a full optimization of elevation will balance the installation cost versus the energy yield for minimizing the LCOE.

### B. Optimal Azimuth Angle (East-West vs. South-North)

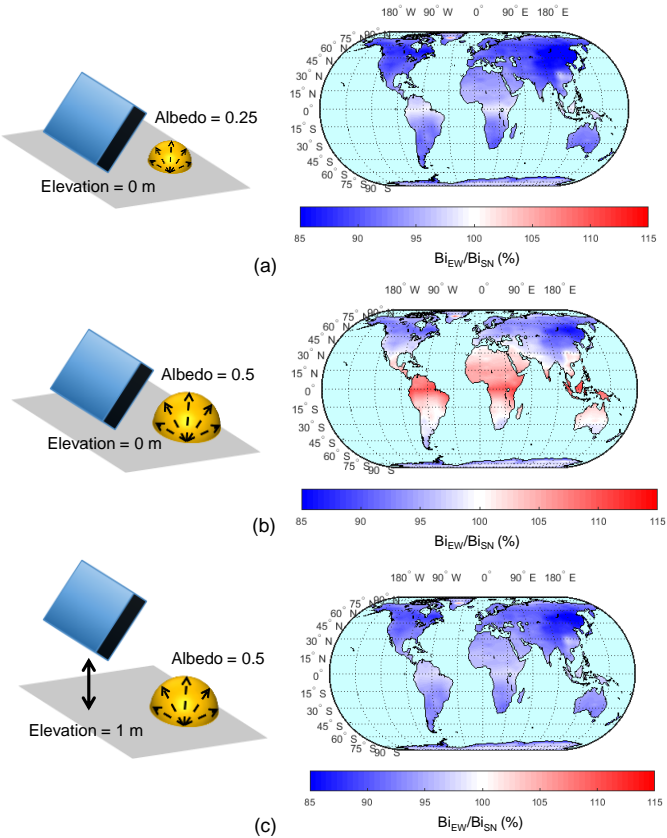

Fig. 7 Global maps showing energy yield ratio of optimally tilted $Bi_{EW}$ over $Bi_{SN}$ for three different scenarios: (a) ground mounted with a ground albedo of 0.25, (b) ground mounted with a ground albedo of 0.5, and (c) 1 m elevated with a ground albedo of 0.5.

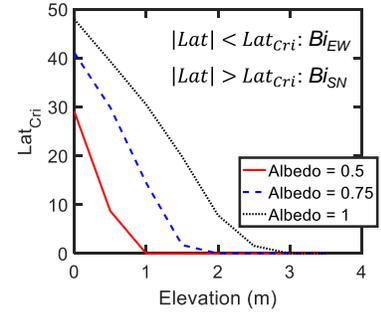

Fig. 8. Critical latitude ($Lat_{Cri}$), below which $Bi_{EW}$ is more favorable than $Bi_{SN}$, and vice versa, as a function elevation for albedo of 0.5, 0.75, and 0.1.

Once the elevation is determined, one must also optimize the orientation of bifacial modules to maximize energy production. Here, we optimize the azimuth angle of bifacial modules at a given elevation. Our simulation reveals that the optimal azimuth angle of bifacial solar modules is essentially either east-west- or south-north-facing, except for the Arctic and Antarctic regions where the bifacial gain is essentially independent of azimuth angle due to the polar day. Therefore, we confine our optimization to two orientations: 1) east-west-facing bifacial modules ($Bi_{EW}$) and 2) south-north-facing bifacial modules ($Bi_{SN}$).

Figure 7 summarizes the performance comparison between $Bi_{EW}$ and $Bi_{SN}$ for the deployment scenarios as presented in Sec. III, i.e., (a) ground mounting with a ground albedo of 0.25, (b) ground mounting with a ground albedo of 0.5, and (c) 1 m elevation with a ground albedo of 0.5. Note that the tilt angles of $Bi_{EW}$ and $Bi_{SN}$ in Fig. 7 are also optimized, which will be discussed in detail later. We point out that across the entire globe, the optimal tilt angle of $Bi_{EW}$ is found to be 90º, i.e., vertical installation, which accords with [17].

**Low Albedo**. Interestingly, our simulation anticipates that $Bi_{SN}$ can outperform vertical $Bi_{EW}$ by up to 15% worldwide for ground mounting with an albedo of 0.25, see Fig. 7(a). With a limited albedo of 0.25, the collection of direct light dictates the total production; vertical $Bi_{EW}$, however, does not absorb any direct light at noon, when direct light peaks. Consequently, $Bi_{SN}$ is more favorable than $Bi_{EW}$ with a low albedo.

**High Albedo**. If the albedo increases to 0.5 at zero elevation, surprisingly, $Bi_{EW}$ can produce more electricity than $Bi_{SN}$ up to 15% within 30º latitude from the Equator, see Fig. 7(b). With albedo equal to 0.5, the contribution of albedo light is comparable to direct and diffuse light. Self-shading of albedo light, however, diminishes the production of $Bi_{SN}$; thus, $Bi_{EW}$ (vertical installation is less susceptible to self-shading) is the preferred configuration. Note that the superior performance of vertical $Bi_{EW}$ shown here has an important implication for bifacial technologies, especially for desert environments (e.g., Saudi Arabia), where $Bi_{EW}$ has the additional advantage of reduced soiling arising from higher tilt angle. Reduced soiling has two advantages, namely, increased energy output and reduced cleaning cost. At higher latitude, the optimal tilt angle $Bi_{SN}$ increases rapidly, which, in turn, diminishes the loss from self-shading. Consequently, $Bi_{SN}$ outperforms $Bi_{EW}$ in regions of high latitude, see Fig. 7(b).

**Elevation.** Remarkably, our simulation indicates that once the modules are mounted more than 1 m above the ground, the optimal orientation of bifacial modules again becomes $Bi_{SN}$ globally, see Fig. 7(c). This change of optimal azimuth angle reflects the fact that elevation reduces self-shading of bifacial modules. Thus, $Bi_{SN}$ suffers less from self-shading and can produce more power than $Bi_{EW}$. As a result, at an elevation of $E_{95}$ with minimal self-shading, the optimum orientation is always south-north facing across the entire world.

**Critical Latitude.** We have shown that $Bi_{EW}$ can outperform $Bi_{SN}$ if self-shading is severe, and vice versa. The magnitude of self-shading at a given location varies as a function of elevation and ground albedo. Specifically, for a given elevation and ground albedo, there exists a critical latitude ($Lat_{Cri}$) below which $Bi_{EW}$ is more productive than $Bi_{SN}$ and vice versa. For example, in Fig. 7(b), $Lat_{Cri}$ is about 30°, with a slight variation along longitude due to the clearness index. Enabled by our simulation framework, we have calculated the average $Lat_{Cri}$ as a function of ground albedo and elevation for different clearness indexes, see Fig. 8. Next, we perform linear regression on our results to develop the empirical equations that calculate $Lat_{Cri}$ based on elevation $E$, module height $H$, and ground albedo $R_A$ as shown below, see Eqns. (A3–A4) in Table A1 of the appendix. These equations will help installers to choose between $Bi_{EW}$ and $Bi_{SN}$ for maximizing electricity yields for a given location and elevation.

### C. Optimal Tilt Angle ($\beta$)

After optimizing azimuth angle, it is important to determine the optimal tilt angle of bifacial modules. As mentioned, for $Bi_{EW}$, vertical installation ($\beta = 90^o$) produces the most electricity. Tilting $Bi_{SN}$ optimally, on the other hand, depends on geographic location and module deployment. Consequently, we have performed a comprehensive study on the optimal tilt angle of $Bi_{SN}$ as a function of latitude, elevation, and albedo, see Fig. 9.

Our simulation results show that the optimal tilt of $Bi_{SN}$ follows the same trend as $Mono_{SN}$ as shown in Fig. 9 (i.e., tilt angle increases with latitude) although the tilt angle of $Bi_{SN}$ is always slightly higher from that of the monofacial counterpart (black dashed lines). This increased tilt enhances the rear-side albedo light collection, consistent with previous studies [11], [14]. The higher tilt angle of $Bi_{SN}$ make them more resistant to soiling compared to monofacial ones, since the soiling loss reduces with increasing tilt angle [60]. Reduced soiling loss will further enhance the bifacial gain of $Bi_{SN}$ relative to $Mono_{SN}$ in the field. Because the optimal tilt angle may differ between $Mono_{SN}$ and $Bi_{SN}$, *the analytical equation previously developed to access optimal tilt angle of monofacial modules is not applicable to bifacial ones.* Therefore, we developed a new set of equations formulated to tilt $Bi_{SN}$ optimally as a function of elevation ($E$), module height ($H$), and ground albedo ($R_A$), whereby we implicitly take the effect of self-shading into account. Equations (A5–A8) are listed in Table A1 of the appendix. The influence of clearness index on optimal tilt is found to be minimal; thus, it has been neglected in these empirical relationships.

Overall, we find that the energy production of bifacial modules optimized by Eqns. (A3–A8) analytically is within 5% relative difference compared to those optimized numerically, which ensures the fidelity of the empirical guidance developed in this paper.

Note that the empirical rules herein are developed for a single standalone bifacial module. At the farm level, in addition to self-shading, a shading effect caused by adjacent rows (i.e., mutual shading) will further diminish the performance, thereby affecting the optimization [41]. For instance, $E_{95}$ is higher for a farm than for a standalone module in order to mitigate mutual shading between each row. We also wish to emphasize the location-specific optimum configuration (Eqs. A1-A8) obtained in this paper assumes an idealized condition (e.g. the absence of shading from nearby objects such as a tree or a chimney, etc.). With these local objects present, a module may have to be tilted/elevated differently from the empirical rule herein. Software tools such as PVsyst [61] that accounts for non-ideal factors (e.g., obstruction shading) should be used in practical design. Obviously, these non-ideal conditions will reduce the energy output on a case-by-case basis.

### V. DISCUSSION AND CONCLUSIONS

In summary, we have developed a comprehensive opto-electro-thermal framework to study and optimize bifacial solar

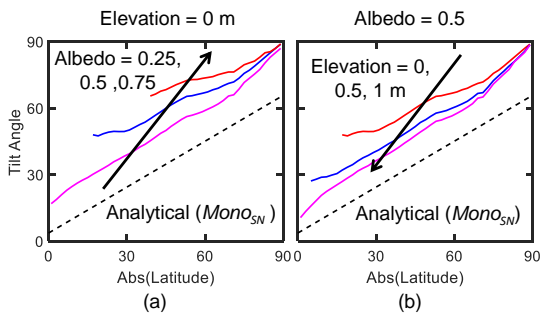

Fig. 9 The optimal tilt angle of $Bi_{SN}$ above $Lat_{cri}$ for (a) albedos of 0.25, 0.5, and 0.75 with ground-mounting and (b) elevations of 0 m, 0.5 m, and 1 m at fixed albedo of 0.5. The optimal tilt angle here is the average over longitudes with different clearness indexes. The arrow indicates the increment of albedo and elevation in (a) and (b), respectively. The black dashed line is the optimal tilt angle for $Mono_{SN}$ obtained analytically from [21].

modules in a global context. The key conclusions of the paper are:
1. Our framework calculates the minute-by-minute solar irradiance data by combining the NASA 22-year average meteorological database [29] with our sophisticated irradiance model for arbitrary location and time. The calculated irradiance is used as inputs into our light-collection model, where the contributions from direct, diffuse, and albedo light are physically and geometrically estimated on both the front and rear surfaces of a bifacial solar module. Here, the effect of self-shading is fully accounted for. Last but not least, we use an opto-electro-thermal coupled framework to self-consistently convert light absorption into annual electricity yield. This framework has also been incorporated into an online simulation tool (Purdue University Bifacial Module Calculator) where a user can model and optimize a bifacial solar module in any arbitrary location in the world [55].
2. Our calculation predicts that for a low ground albedo of 0.25 corresponding to vegetation/soil, ground-mounted bifacial solar modules can only achieve bifacial gain up to 10% relative to their monofacial counterpart across the entire world (except for the Arctic and Antarctic regions). However, by boosting the albedo to 0.5 via artificial ground reflectors as well as lifting modules 1 m above the ground surface to reduce self-shading, one can potentially enhance the bifacial gain up to 30%. Hence, our finding encourages the future development of cost-effective ground reflectors and module-elevating schemes to make bifacial modules more financially viable.
3. We demonstrate the enormous impact of self-shading on the optimization of bifacial solar modules. Our analysis reveals that under severe self-shading, i.e., high albedo and low elevation, the vertical $Bi_{EW}$ configuration is superior because $Bi_{SN}$ is more prone to self-shaded albedo loss. For instance, for an albedo of 0.5 and zero elevation, vertical $Bi_{EW}$ can outperform $Bi_{SN}$ up to 15% below the latitude of 30º, and vice versa beyond the latitude of 30º. In contrast, with a reduced albedo to 0.25, i.e., less self-shading, $Bi_{SN}$ is more beneficial than $Bi_{EW}$ across the globe.
4. Enabled by our rigorous simulation framework, we have developed a set of empirical design rules to analytically and optimally configure bifacial solar modules in arbitrary geographic locations. Specifically, they can 1) determine the minimum elevation to achieve 95% of the maximum self-shading-free energy production, above which further elevating the modules will offer insufficient benefits, 2) locate the critical latitude $Lat_{Cri}$ below which an east-west orientation is more favorable than the south-north orientation, and 3) calculate the optimal tilt angle of bifacial modules. These empirical equations (within 5% relative difference compared to numercial simulation) enable rapid design of bifacial modules globally without performing sophisticated local optimization.

Finally, there are three additional aspects that are beyond the scope of this paper but are still important regarding the economic viability of the bifacial technology. First, our paper emphasizes the optimization of a standalone bifacial module related to module cost, *whereas a farm-level optimization is equally crucial to reducing other cost associated with land usage* [38]. At the farm scale, mutual shading between each row of solar panels can curtail the total energy production, which requires future investigation [62]. In a future farm-level study, the economic tradeoff between module and land cost must be balanced carefully.

Second, long-term in-field reliability of the solar modules must be considered to calculate LCOE of bifacial solar modules [63]—a topic not discussed in this paper. For example, it has been demonstrated experimentally that compared to conventional tilting, vertical installations is immune to soiling degradation; so cleaning costs and water usage are significantly reduced [64]. In practice, these factors ought to be fully analyzed for LCOE calculation.

Last but not least, the analysis herein adapts the NASA satellite-derived insolation database. Despite being suitable for global calculation (comprehensive geographic coverage), the satellite-derived insolation data is expected to be less accurate than that measured by the ground-level weather station (rRMSE = 10.25%) [65]. As a result, the analytical design guidance developed in this paper can accelerate the design cycle for bifacial solar modules as a preliminary estimation. A much more careful local construction optimization based on detailed local meteorological database [66] must be carried out for the actual installation site [61], [67] where local non-idealities (e.g., obstruction shading, finite ground size) must be carefully and comprehensively examined to ensure the financial viability of bifacial technologies.

# APPENDIX

*Lat*: Latitude

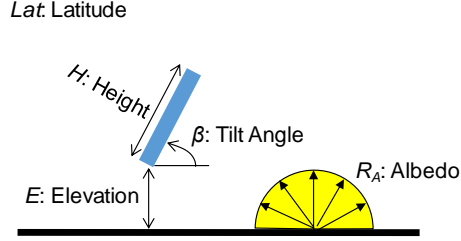

Fig. A1  Physical definitions of the parameters in Table A1

Table A1. A set of analytical equations to optimize the elevation and orientation of bifacial solar modules

| $E_{95}$ in meter for a module height of $H$ | | |
|---|---|---|
| $E_o = H \times (-Lat \times (0.028 \times R_A + 0.009) + 3.3 \times R_A + 0.4)$ | (A1) | $E_{95}$ is the minimum elevation to achieve at least 95% of the self-shading absent maximum energy yield, i.e., further elevation only provides limited energy boost. |
| $If\ E_o \leq 0,\ E_{95} = 0\ and\ If\ E_o > 0, E_{95} = E_o$ | (A2) | |
| $Lat_{Cri}$ of bifacial solar module for a given elevation ($E$), module height ($H$), and albedo ($R_A$) | | |
| $Lat_o = E/H \times (44 \times R_A - 62) + 37 \times R_A + 12$ | (A3) | $Lat_{Cri}$ is the critical latitude below which $Bi_{EW}$ produces more electricity than $Bi_{SN}$ and vice versa. |
| $If\ Lat_o \leq 0,\ Lat_{Cri} = 0^o\ and\ If\ Lat_o > 0,\ Lat_{Cri} = Lat_o$ | (A4) | |
| Optimal Tilt Angle $\beta_{Opt}$ for $Bi_{SN}$ for a given latitude ($Lat$), elevation ($E$), module height ($H$), and albedo ($R_A$) | | |
| $\beta_o = a \times Lat + b$ | (A5) | $\beta_{Opt}$ is the optimal tilt angle for $Bi_{SN}$ for maximum electricity yield |
| $a = 0.86 - 0.57 \times R_A \times \exp(-E/H)$ | (A6) | |
| $b = 4.5 + 62 \times R_A \times \exp(-E/H)$ | (A7) | |
| $If\ \beta_o \geq 90^o,\ \beta_{Opt} = 90^o\ and\ If\ \beta_o < 90^o,\ \beta_{Opt} = \beta_o$ | (A8) | |

# Optimization and Performance of Bifacial Solar Modules: A Global Perspective

Xingshu Sun,[1] Mohammad Ryyan Khan,[1] Chris Deline,[2] and Muhammad Ashraful Alam[1]

[1] *Network of Photovoltaic Technology, Purdue University, West Lafayette, IN, 47907, USA.*

[2] *National Renewable Energy Laboratory, Golden, Colorado, 80401, USA*

# Supplementary Information

Here we will present four tables of global maps summarizing the optimization and performance (i.e., tilt angle, azimuth angle, annual energy yield, and bifacial gain) for a bifacial solar module with different deployment scenarios (i.e., elevation and albedo).

Table SI1. Optimal tilt angle of a 1 m high module for different ground albedo and elevations (E)

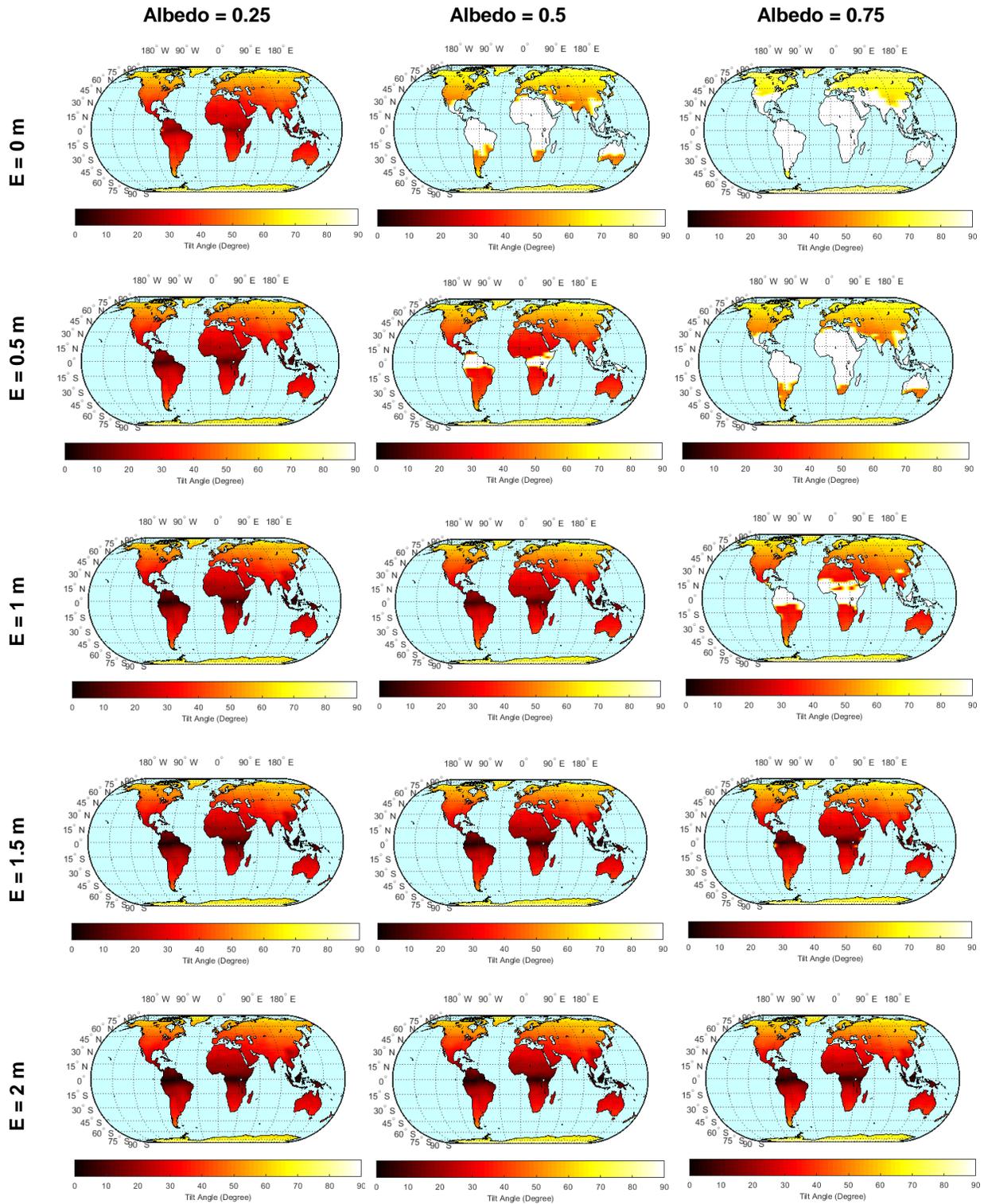

Table SI2. Optimal azimuth angle of a 1 m high module for different ground albedo and elevations (E)

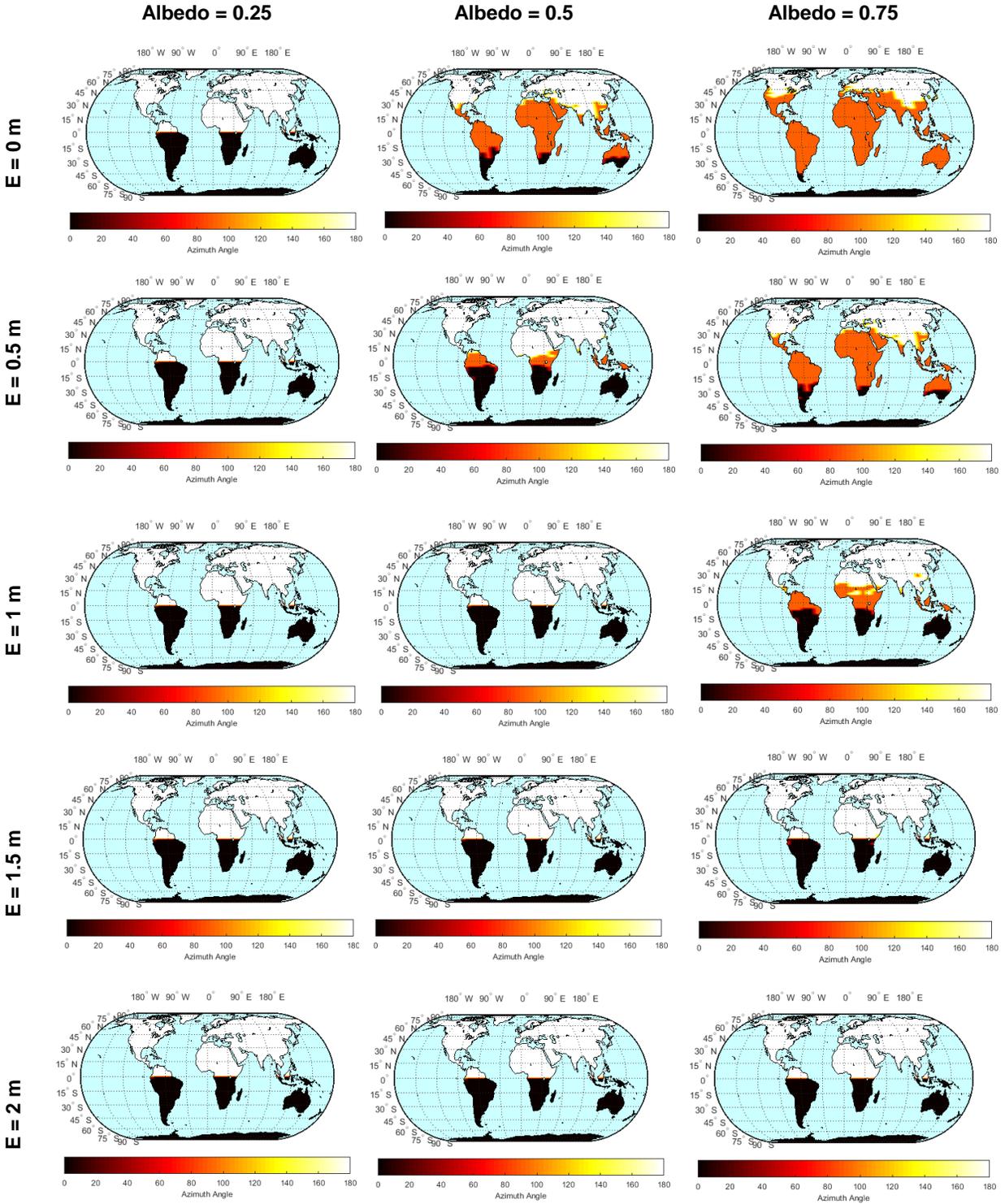

Table SI3. Maximum annual electricity yield of a 1 m high module for different ground albedo and elevations (E)

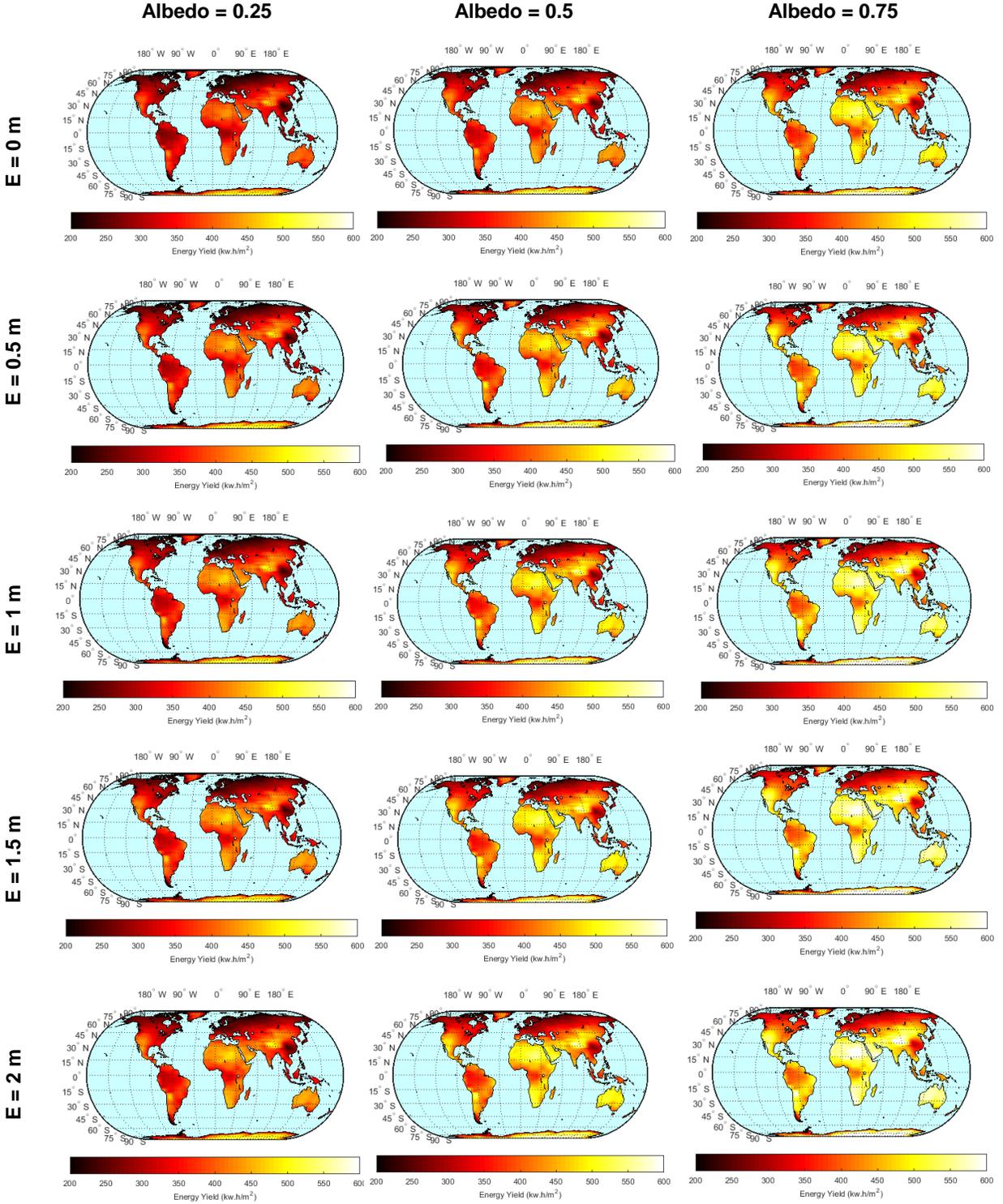

Table SI4. Maximum bifacial gain of a 1 m high module for different ground albedo and elevations (E)

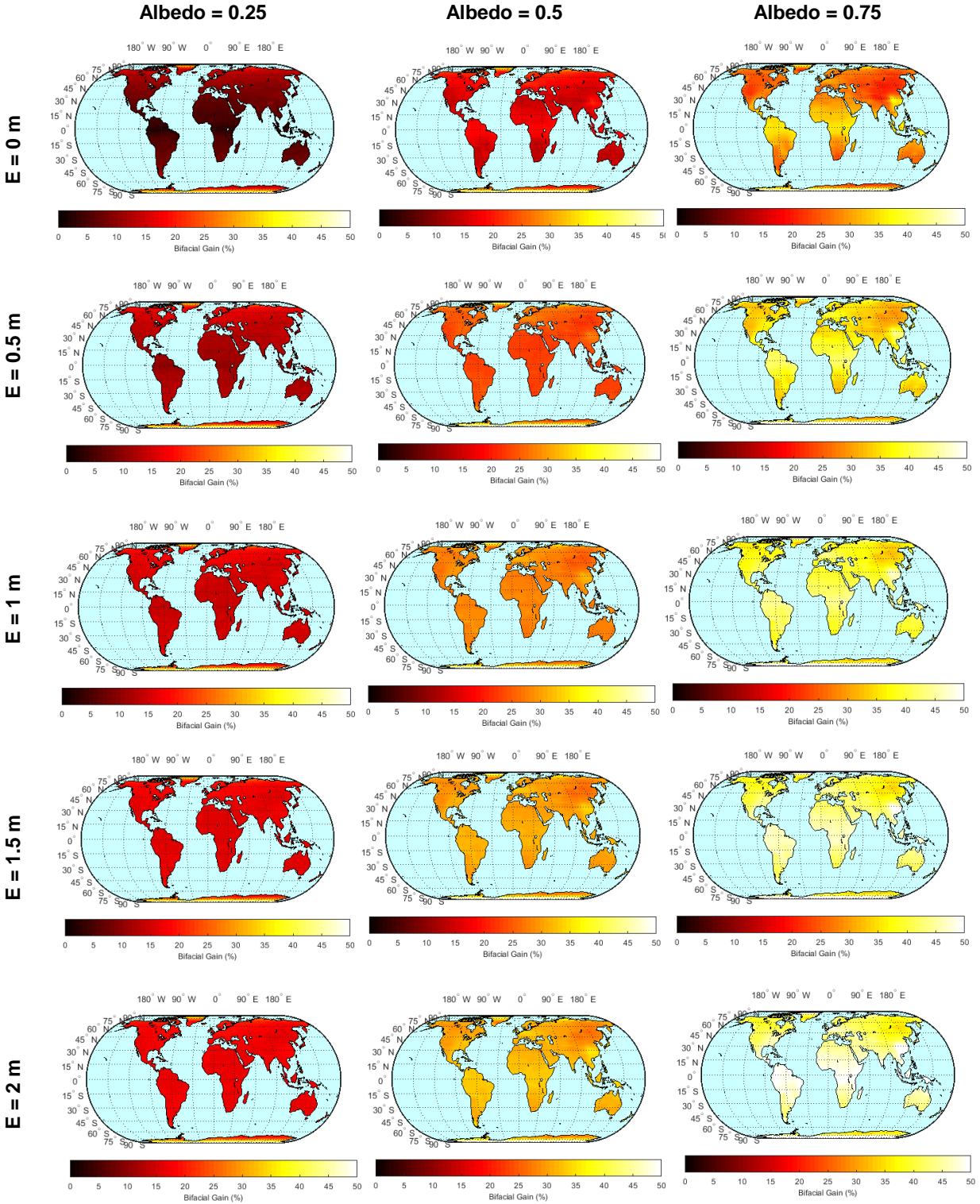